# Growth of and optical emission from GaMnAs thin films grown by molecular beam epitaxy

J. F. Xu, S. W. Liu, Min Xiao, and P. M. Thibado[a]

*Department of Physics, University of Arkansas, Fayetteville, Arkansas 72701*

GaMnAs thin films with different Mn doping concentrations were grown via molecular beam epitaxy using a substrate temperature of 250 °C. The thin films were investigated using photoluminescence (PL) measurements from 8 to 300 K. Transitions involving Mn acceptors were identified and a binding energy of ~0.1 eV was found. A Mn doping concentration dependent PL spectrum was found to lend insight into the film quality at a local level. Temperature dependent PL studies show that the doping related emissions drop faster in energy than other peaks with increasing temperature, indicating that they are more sensitive to changes in the surrounding environment.

[a] Electronic mail: thibado@uark.edu

## I. INTRODUCTION

The incorporation of transition metal dopants with semiconductors may open up opportunities for utilizing a carrier's spin property in conventional electronic devices.[1,2] The Mn-doped GaAs (GaMnAs) system grown by molecular beam epitaxy (MBE) has attracted much attention in this area of study.[3,4] However, the Curie temperature $T_C$ of GaMnAs has been limited to ~150 K at this moment.[5] The reason for suppression in $T_C$ has been derived from the extremely low Mn solubility limit in GaAs (Ref. 6) and high defect density associated with the Mn doping. At present, some of the best samples of GaMnAs have been grown by MBE and using a low-substrate temperature.[7] The low-temperature MBE growth allows one to dope GaAs with Mn beyond its solubility limit, making it possible to realize a III-V-based diluted magnetic semiconductor. However, the low-temperature growth also induces more defects in GaMnAs. It is important to investigate and optimize the sample growth process so that the GaMnAs system has a relatively high Mn solubility and low defect concentration.

In GaMnAs, when Mn substitutes at Ga sites it will take on the dual role of acceptor and local magnetic moment.[8,9] In this article, we have grown GaMnAs thin films with different levels of Mn doping by MBE at a substrate temperature of 250 °C. The photoluminescence (PL) spectra of these GaMnAs films were investigated using a temperature range of 8–300 K. Mn concentration dependent PL spectra from GaMnAs thin films were also obtained, and these provide information about how the local film quality changes with doping concentration.

## II. EXPERIMENTAL PROCEDURE

Samples were prepared in an ultrahigh vacuum (UHV) MBE growth chamber (Riber 32) having a base pressure of $\sim 2 \times 10^{-10}$ Torr. The MBE chamber includes Ga and Mn effusion cells together with a two-zone As valved-cracker cell. It is also equipped with a reflection high-energy electron diffraction (RHEED) system. The highest-quality commercially available, "epiready," semi-insulating 2 in. diameter GaAs(001)±0.1° wafers (AXT, Inc., etch pit density <5000 cm$^{-2}$, full width at half maximum=3.8 arc sec) were used in this study after being cleaved into quarters. One quarter was mounted on a 2 in. diameter standard MBE molybdenum block using indium as solder. The substrate was then loaded into the load-lock chamber without any chemical cleaning. Next, the substrate was transferred to the heating stage inside the MBE chamber, and the chamber was cooled down using liquid nitrogen. The substrate was heated to 580 °C while exposing the surface to As$_4$ to remove the surface oxide layer. A thin buffer layer of GaAs was grown on the substrate for 5 min. During this time RHEED oscillations were used to determine that the growth rate of the GaAs was 780 nm/h. Next, the substrate temperature ($T_S$) was set to the desired growth temperature of 250 °C. GaMnAs films were then grown for 1 h, while RHEED was used to monitor the surface reconstruction during and following the growth. After growth, the sample was cooled down and removed from the UHV system. Samples were cleaved into multiple smaller pieces (5×5 mm$^2$) for characterization measurements.

The PL measurements were performed in a variable temperature (8–300 K) closed-cycle helium cryostat. The 532 nm line from a double neodymium-doped yttrium aluminum garnet laser was used for continuous-wave PL excitation. The PL signal from the sample was dispersed by a monochromator (resolution ~0.001 eV) and detected by a liquid-nitrogen-cooled charged-coupled device.

## III. RESULTS

For Mn-doped GaMnAs films, the thickness range is between 0.6 and 1 μm. The Mn distribution within each sample is uniform as determined by Auger electron spectroscopy or secondary ion mass spectrometry depth profiling.[10] The Mn concentration in GaMnAs films depends on the Mn cell temperature. The relation between Mn concentration in GaMnAs and Mn cell temperature is shown in Fig. 1. The Mn concentration increases exponentially with increasing Mn cell temperature, as expected. For a modest cell temperature range from 700 to 900 °C the Mn concentration has a very large dynamic range from $\sim 10^{18}$ to $\sim 10^{22}$ cm$^{-3}$. The PL spectra of the GaMnAs films grown using various Mn cell temperatures are shown in Fig. 2 (the Mn cell temperature is labeled at the left edge of each line plot). The spectra for these samples were taken at 8 K. At a Mn cell temperature of 700 °C, four luminescence peaks can be identified, namely, the GaAs exciton line (labeled a) at 1.506 eV, an impurity-induced recombination peak (labeled b) at 1.487 eV, and deeper bands (labeled d and e) near 1.369 and 1.452 eV, respectively. At a Mn cell temperature of 750 °C, the luminescence spectrum is similar to that grown at Mn cell temperature of 700 °C. As the Mn concentration in the sample increases further (Mn cell temperature of 800 °C), a new PL peak at 1.404 eV (labeled c) appears. However, the new PL peak disappears as the Mn concentration continues to increase (cell temperature of 850 °C). At a Mn cell temperature of 900 °C, the GaAs exciton peak also disappears.

The temperature dependence of the PL spectrum for the GaMnAs film grown using a Mn cell temperature of 800 °C was investigated. The relation between the sample temperature and the PL peak energy position and its intensity for the four strongest peaks are shown in Figs. 3(a) and 3(b), respectively. All emission peaks shift to lower energy when the temperature increases, and all the peak intensities decrease with increasing temperature.

## IV. DISCUSSION

We begin our discussion by first explaining the origin of the various peaks shown in Fig. 2. Three of the PL peaks can be easily identified from the literature. According to the location and width, the peak at 1.506 eV (labeled a) should be the GaAs exciton radiative transition, as mentioned earlier. Given that the samples are grown at 250 °C it is possible that some or all of the intensity in this peak is coming from the bulk substrate. The peak at 1.487 eV (labeled b) is also well known as the donor-acceptor pair transition involving a carbon at an As site.[11] Note that we have acquired PL data from the bare substrate, and we see a very strong GaAs exciton peak and a weaker carbon peak. The emission band around 1.369 eV (labeled d) is thought to be the longitudinal-optical (LO) phonon mode in GaMnAs.[12] No peaks are observed from the bare substrate where the Mn related peak is found. The PL emission peak at 1.452 eV (labeled e) has never been reported before in GaMnAs or GaAs. This peak may be a Mn related emission; however, further studies are needed.

The PL peak at 1.404 eV (labeled c) only appears in the sample grown at a Mn cell temperature of 800 °C (see Fig. 2). This peak is associated with the transition from the conduction band to the Mn acceptor level.[13] This peak is a direct confirmation that Mn atoms are occupying Ga sites. In fact, this Mn related peak gives rise to a calculated acceptor binding energy of 102 meV above the valence band edge. This is in good agreement with values obtained by many different methods.[14,15]

It is surprising that the peak at 1.404 eV (labeled c) is only observed in the film grown at a Mn cell temperature of 800 °C. At lower Mn cell temperatures (700 or 750 °C), the Mn atom concentration as well as Mn acceptor concentration in the thin films is low, so the emission related to Mn acceptors is hard to detect. However, when the Mn doping level is higher (e.g., at Mn cell temperature of 850 °C), the peak disappears again. At even higher Mn concentration levels (e.g., at Mn cell temperature of 900 °C) the GaAs exciton peak even disappears. This suggests that the high doping induces so many defects into the system that nonradiative decay channels may dominate. However, we cannot rule out that the film is now opaque enough to block the substrate exciton peak.

Let us now turn our attention to the temperature dependence of the PL peak energy shown in Fig. 3(a). When the sample temperature increases, the emission peak energy shifts to the lower-energy side due primarily to lattice dilation. In order to quantify the role of this shift we model the shift next. The temperature dependence of the peak positions can be expressed by the Varshni formula,[16]

$$E(T) = E(0) - \beta T^2/(T + \gamma), \qquad (1)$$

where $E(0)$ is the gap energy at 0 K. $T$ is the absolute temperature. The parameter β is the band gap slope in the high temperature limit, and γ is a fitting parameter. From the fits, we obtained the parameters listed in Table I. Notice that the β value for Mn acceptor related peak is $2.2 \times 10^{-4}$ eV/K, which is larger than the other three peaks. This indicates that the emission peak energy position shifts faster with increasing temperature compared to other peaks. So, the lattice dilation plays a more important role at the Mn donor sites.

The γ value represents a low-temperature fitting parameter. We report the values here, but it is more difficult to interpret its meaning. The γ values for GaAs and Mn related peaks are smaller than those of other two peaks.

The intensity of all four peaks also decreases with temperature [see Fig. 3(b)]. The intensity decreases fastest for the two peaks involving doping [C at As (shaded circle symbols) and Mn at Ga (shaded square symbols)]. Peak intensity drops with increasing temperature primarily due to the thermally activated electron-phonon interaction.[17] This drop in intensity indicates that the doped atoms have different properties from the host ones. This difference is likely due to variations induced in the local lattice structure.

## V. SUMMARY

Mn-doped GaAs (GaMnAs) thin films were grown on semi-insulating GaAs(001) substrates using MBE at a substrate temperature of 250 °C for a very broad range of Mn doping concentrations. The PL spectra show that low-doped and high-doped samples do not have significant Mn on Ga sites, while medium-doped samples clearly show Mn on Ga sites with a binding energy of ~0.1 eV. The emission peaks involving doping (Mn at Ga site and C at As sites) drop faster with temperature than others, indicating that the doped atoms are more sensitive to changes in the surrounding environment.

## ACKNOWLEDGMENT

The authors would like to acknowledge the support for this work from National Science Foundation under Grant No. DMR-0405036.

FIG. 1. Mn concentration in Mn-doped GaAs thin films grown using a substrate temperature of 250 °C vs Mn cell temperature. The concentration was determined using Auger electron spectroscopy and secondary ion mass spectroscopy depth profiling.

FIG. 2. PL spectra acquired at 8 K for various samples prepared using different Mn cell temperatures (labeled at the left of each line plot). The PL peaks are labeled as follows: (a) GaAs exciton radiative transition (possibly coming from the substrate), (b) carbon acceptor related emission, (c) Mn acceptor related emission, (d) LO optical phonon related peak, and (e) possibly another Mn related peak.

FIG. 3. (a) PL peak energy position as a function of the sample's temperature. (b) PL peak intensity as a function of the sample's temperature. The PL spectra are taken from the Mn-doped GaAs thin film grown using a Mn cell temperature of 800 °C: (open square) GaAs exciton radiative transition peak, (filled circle) carbon acceptor related peak, (filled square) Mn acceptor related peak, and (open triangle) LO phonon peak.

TABLE I. Varshni formula fitting parameters calculated by fitting the experimental data shown in Fig. 3. The column headings are the peak energy at 8 K $E(8)$, the peak energy at 0 K $E(0)$, and then the parameters $\beta$ and $\gamma$ [see formula (1)].

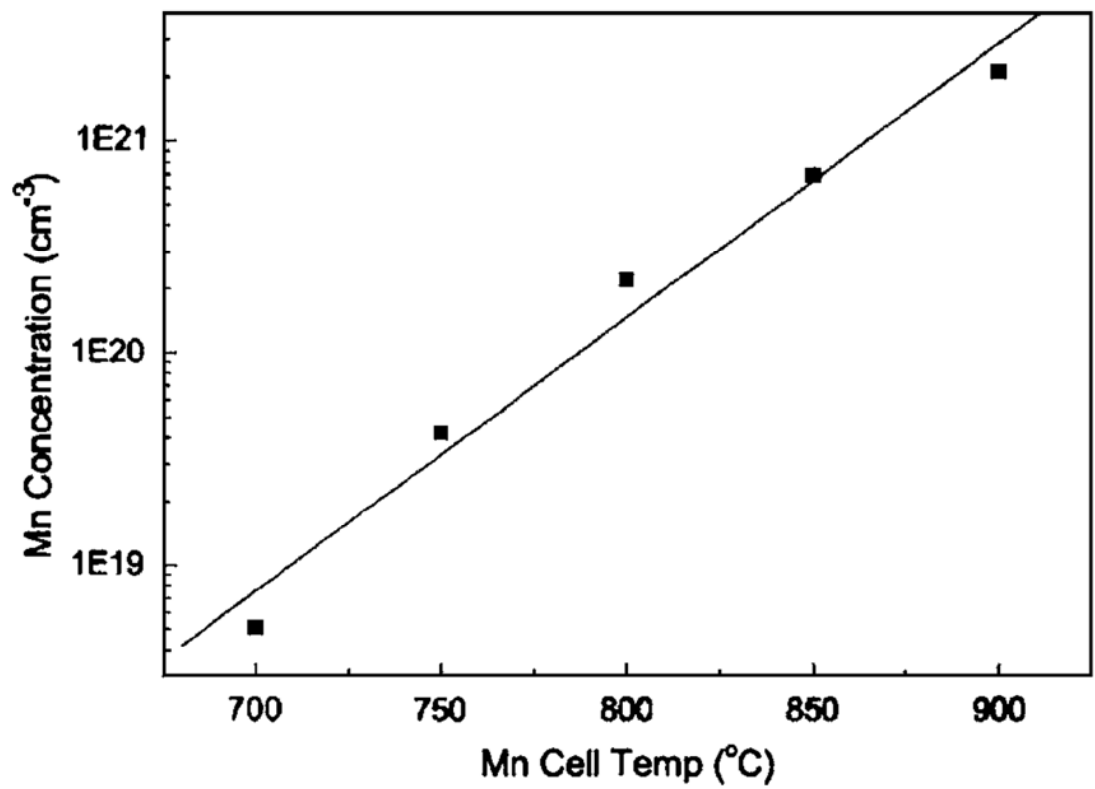

Figure 1.

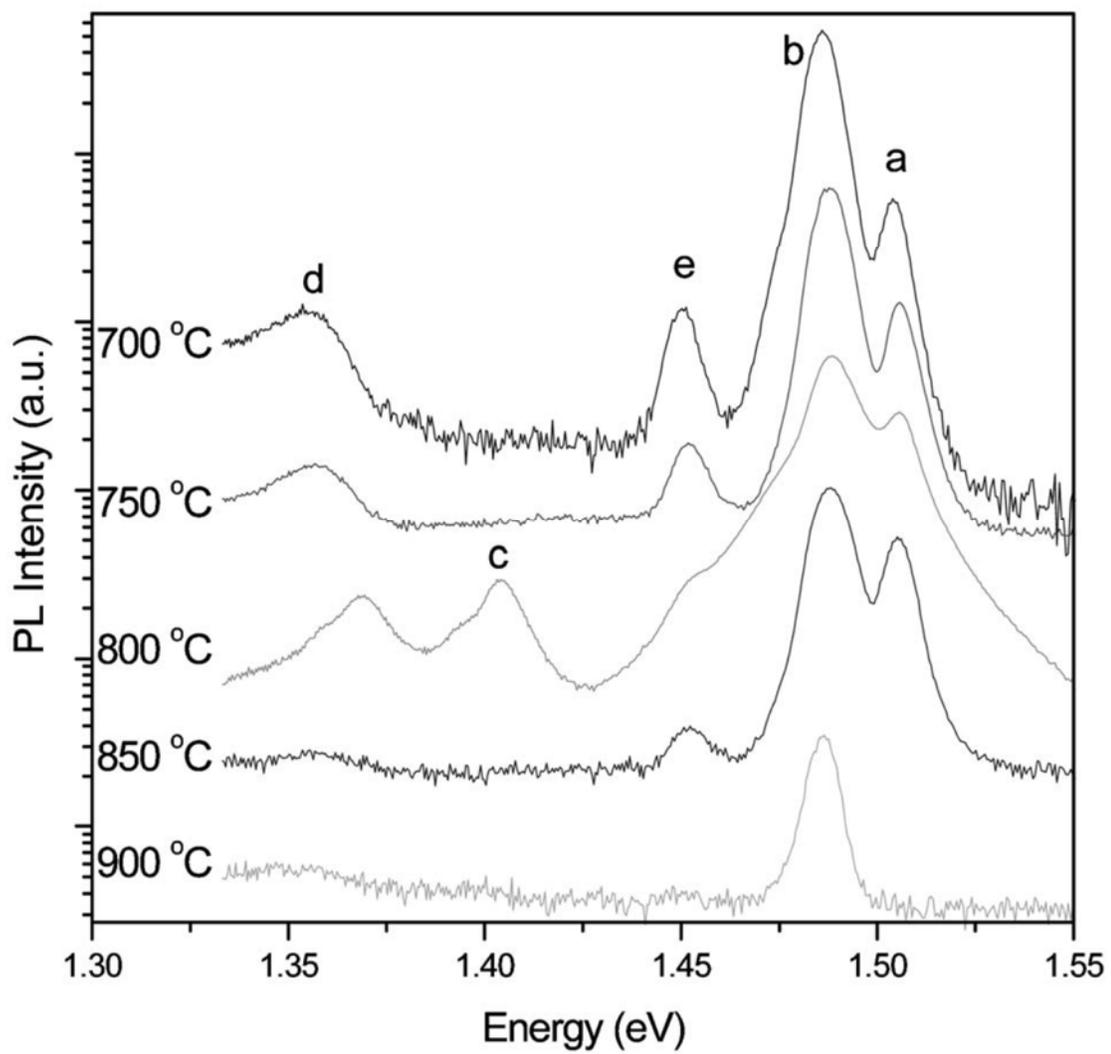

Figure 2.

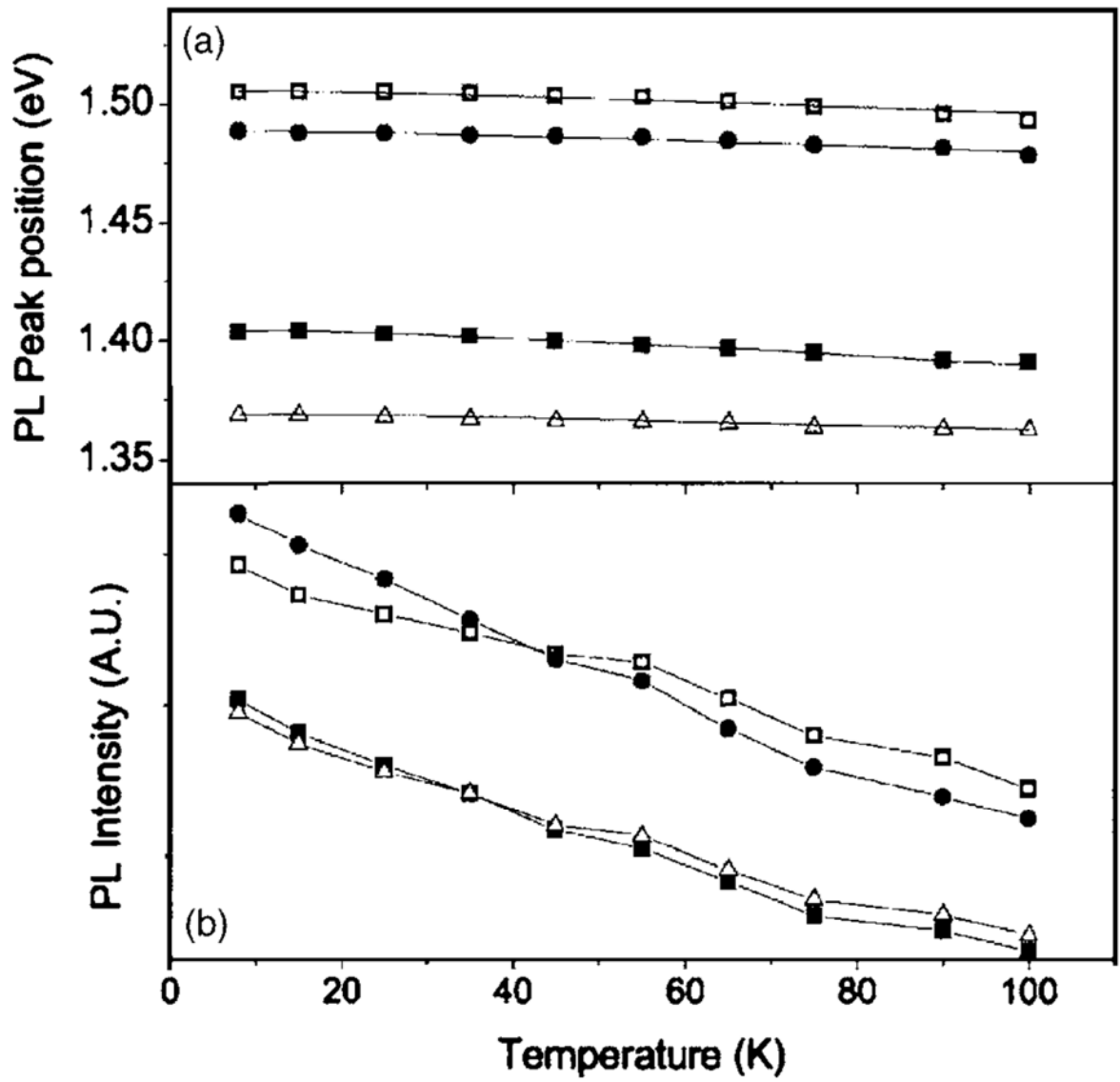

Figure 3.

| $E(T=8\text{ K})$ (eV) | $E(T=0\text{ K})$ (eV) | $\beta$ ($10^{-4}$ eV K$^{-1}$) | $\gamma$ (K) |
| --- | --- | --- | --- |
| 1.506 | 1.506 | 1.4 | 45 |
| 1.487 | 1.489 | 1.5 | 65 |
| 1.404 | 1.405 | 2.2 | 44 |
| 1.369 | 1.369 | 1.3 | 89 |

Table 1.